\documentclass[12pt,preprint]{aastex}

\slugcomment{}

\shorttitle{Radio Bursts on AD Leo}
\shortauthors{Osten \& Bastian } 

\begin{document}
 
\title{Ultra-high-time Resolution Observations of Radio Bursts  on AD~Leonis }
 
\author{Rachel A. Osten\altaffilmark{1}}
\affil{Astronomy Department, University of Maryland, College Park MD 20742 \\
Electronic Mail: rosten@astro.umd.edu}
\author{T. S. Bastian}
\affil{National Radio Astronomy Observatory, 520 Edgemont Road, Charlottesville, VA 22903 \\
Electronic Mail: tbastian@nrao.edu}
\altaffiltext{1}{Hubble Fellow}

\begin{abstract}
We report observations of a radio burst that occurred on the flare star AD~Leonis 
over a frequency range of 1120-1620~MHz ($\lambda\approx$18--27~cm). These observations, 
made by the 305~m telescope of the Arecibo Observatory, are unique in providing the highest 
time resolution (1~ms) and broadest spectral coverage ($\Delta \nu/\nu=0.36$) of a stellar radio 
burst yet obtained. The burst was observed on 2005 April 9. It produced a peak flux density of $\sim 500$ 
mJy and it was essentially 100\% right-circularly polarized. The dynamic spectrum shows a rich 
variety of structure: patchy emission, diffuse bands, and narrowband, fast-drift striae. Focusing 
our attention on the fast-drift striae, we consider the possible role of dispersion and find that it 
requires rather special conditions in the source to be a significant factor. We suggest that the 
emission may be due to the cyclotron maser instability, a mechanism known to occur in planetary 
magnetospheres. We briefly explore possible implications of this possibility.  
\end{abstract}


\section{Introduction}

The use of radio dynamic spectra has played a central role in identifying and clarifying the 
physical mechanisms at work in the solar corona \citep[see][for reviews]{1985srph}.  The 
application of similar techniques to active stars has long been an important goal, but it has been 
hampered by limitations in available instrumentation. Past studies of radio emission from M dwarf 
flare stars led to the discovery of extreme stellar radio bursts, characterized by close to 100\% 
circularly polarized emission with brightness temperatures in excess of 10$^{14}$K and durations 
less than a few tens of milliseconds \citep{gudel1989,bastian1990}.
However, these 
spectroscopic investigations of the coherent radio bursts on flare stars have typically been limited 
by relatively long integration times \citep{bb1987,gudel1989} and/or limited frequency bandwidth 
ratios $\Delta\nu/\nu$  \citep[usually just a few percent; e.g.,][]{bastian1990, abada1997a}. The 
necessary combination of high time resolution and a large frequency bandwidth ratio has only been 
available infrequently \citep{stepanov2001, zaitsev2004}, precluding measurements of key parameters 
such as the intrinsic frequency bandwidth or frequency drift rate of the radio bursts, making the 
interpretation of these puzzling events difficult.  It is only with the recent advent of radio 
spectrometers capable of supporting both a large bandwidth ratio and high time resolution simultaneously 
that progress in understanding the physics of radio bursts in the coronas of other stars becomes possible. 

AD~Leonis, a young disk star at a distance of 4.9 pc from the Sun, is one of the most active flare stars 
known, producing intense, quasi-steady chromospheric and coronal emissions 
\citep{hawleyetal2003, hunschetal1999, jackson1989} seen at UV, X-ray, and radio wavelengths. The 
star is also highly variable, producing flares from radio to X-ray wavelengths  
\citep[e.g.,][]{bastian1990, hawleypettersen1991, hawleyetal2003, hawleyetal1995, favataetal2000}. 
Its propensity for frequent and extreme radio bursts 
\citep[with intensities peaking at $>$ 500 times the 
quiescent radio luminosity of 5.5$\times$10$^{13}$ erg s$^{-1}$ Hz$^{-1}$;][]{jackson1989} makes 
it a frequent target for radio investigations of stellar flares. In a previous paper 
\citep[][hereafter, Paper I]{ob2006} 
we described the initiation of a pilot program to observe active M dwarfs with the Arecibo 
Observatory's Wideband Arecibo Pulsar Processor (WAPP), and first results from that program.  
Here we describe the next phase, which increased the time resolution by a factor of 10 to 1~ms.

\section{Observations \label{analsec}}

We observed AD~Leo with the 305~m telescope at Arecibo Observatory\footnote{The Arecibo Observatory is part of the
National Astronomy and Ionosphere Center, which is operated by Cornell University under a cooperative agreement
with the National Science Foundation.} on each of four days from 2005 April 8--11, 
during which time approximately $4\times 4$ hours of data were collected. The observations were made 
using the ``L-band wide" dual-linear feed and receiver (1100-1700 MHz) with the Wideband Arecibo Pulsar 
Processor (WAPP) backend.  The WAPP was selected because it provides the means of observing a large 
instantaneous bandwidth with excellent spectral and temporal resolution. The WAPP provides four data 
channels, each of 100~MHz bandwidth.  These were deployed across the L-band wide receiver as follows: 
1120--1220 MHz, 1320--1420 MHz, 1420--1520 MHz, and 1520--1620 MHz.  A gap was deliberately left between 
1220--1320 MHz to avoid the strong radio frequency interference (RFI) present in this frequency range.  
Nevertheless, the effects of other sources of RFI could not be entirely avoided in the frequency bands 
observed. We employed a data acquisition mode where data were sampled with a time resolution of 1~ms; 128 
spectral channels were sampled across each 100 MHz channel, yielding a spectral resolution of 0.78 MHz. All 
four correlation products were recorded between the native-linear X and Y feed elements (XX, YY, XY, YX) 
with three-level sampling.  Our observing strategy was to observe the target, AD Leo, for 10 minute scans, and 
then to inject a correlated calibration signal to determine the amplitude and phase calibration of the two 
polarization channels. A standard calibrator source, B1040+123, was also observed prior to 
each of the four observing runs.

The beam size using the L-band wide receiver at 1.4 GHz is $\approx3.1\times$3.5 arcminutes in 
azimuth and zenith angle. Interferometric measurements of AD~Leo's 1.4~GHz flux density during 
periods of apparent quiescence are $\approx2$ mJy \citep{jackson1989}, well below the confusion 
limit of Arecibo at the observed frequencies ($\approx$ 16--38 mJy from 1120 MHz to 1620 MHz).  Although 
there is a bright background radio source located $\approx$ 2.3 arcminutes away from AD~Leo 
\citep{seiradakis1995} and therefore within the primary beam, we utilized a ``time-switching'' 
scheme to difference times of burst activity (``on") and the quiescent state plus background (``off"). The 
source of radio bursts has been confirmed as AD~Leo in past years using a variety of techniques: 
interferometric observations by the VLA  \citep{bb1987}, or the characteristic 
frequency modulation of sources in the main beam of the line feeds formerly used at Arecibo 
\citep{bastian1990}.
Our calibration approach enabled us to estimate the antenna temperature and flux density through the 
use of gain curves as a function of azimuth and zenith angle. We determined the rms uncertainty in 
regions of the dynamic spectrum where no bursting behavior was evident, estimating the 1 $\sigma$ 
noise uncertainty for various time resolutions to be $(1.6-2.2)/\sqrt \tau$ mJy/channel, where 
$\tau$ is the integration time in seconds. 

\section{Data Analysis and Results}

A spectacular radio burst occurred on 2005 April 9 at approximately 01:15:40 UT. An overview of the 
event is shown in Fig.~1 with a time resolution of 100~ms. Note the data gap between 1220-1320 MHz and 
horizontal features due to RFI. The radio burst is essentially 100\% right-circularly polarized 
(RCP; Fig.~1a). The emission is bounded in frequency from above, the upper limit ranging from $\approx 
1200-1600$ MHz. The low frequency limit of the emission is unknown because of a lower limit imposed by the 
spectrometer. The emission presumably extends well below 1120~MHz.

While the burst was only $\approx45$~s in duration, it was observed with a time resolution of 1~ms. 
In the following, we confine our attention to only a small part of the data which nevertheless shows a 
rich variety of spectral signatures. The horizontal bar shown in Fig.~1, spanning the most intense RCP 
emission, indicates the 10~s interval shown in Fig.~2 with a time resolution of 10~ms. The four consecutive 2~s time intervals
labeled (A-D) in Fig.~2 are, in turn, shown with 1~ms time resolution in Fig.~3. It is readily apparent 
that the radio emission is spectrally and temporally complex.  Details of time intervals a--d$_{2}$ are 
shown in Figs. 4-8 over restricted bands but with 1~ms time resolution.
 
Fig.~4 is of particular interest. It shows a 0.75~s interval of the spectrum corresponding to time 
interval ``a" in Fig.~3, panel A. The frequency interval shown is 1120-1220~MHz. An intense series of striae are 
visible, drifting from high to low frequencies with time. 
It is difficult to determine whether an underlying diffuse component exists or whether the striae 
recur so fast that they overlap and merge. It is therefore difficult to isolate and characterize the 
bandwidths and durations of single stria.  Nevertheless, their durations have been constrained in aggregate 
as follows: first, the spectrum was smoothed in time channel by channel using a 50~ms running mean. Second, the smoothed spectrum was subtracted from the observed spectrum yielding a residual spectrum containing the rapidly varying striae. Third, to improve the signal to noise ratio, the residual spectrum was then ``corrected" for the frequency drift of the striae by progressively shifting each channel in time to yield vertical striae. Note that if the drift rate were due to group delay, this operation would be referred to as de-dispersion; however, see 
\S4.1. It is found that a range of drift rates is present in the striae, corresponding to 
delays of 43-48~ms per 100 MHz. The mean delay is 45~ms per 100 MHz.  Therefore the mean drift 
rate is -2.2 GHz s$^{-1}$ with a range of -2.33 to -2.08 GHz s$^{-1}$ observed. Finally, the 
drift-corrected, residual spectrum was frequency-averaged over the central seventy (RFI-free) 
channels. All striae with an SNR$>4$ were accumulated, 32 in all. These have a mean FWHM duration 
of $\delta t\approx 2$ ms. A limit on the intrinsic source size $d$ follows by noting that 
$d<c\ \delta t\approx 600$~km.  Coupled with the frequency drift rate of the striae, the mean 
duration implies a mean instantaneous bandwidth of 4.4 MHz, or $\Delta\nu/\nu\lesssim 0.5\%$. The 
recurrence rate of the striae is $>40$ s$^{-1}$ in this example.

Turning to other examples, Fig.~5 also shows 0.75~s of data (interval ``b" in Fig.~3, panel B) from 1120-1220 MHz. The emission at this time, only 2~s after interval ``a", is far more diffuse than that shown in Fig.~4 
although fast-drift structure is still visible.  Fig.~6 shows 0.5~s of data (interval ``c" in Fig.~3, panel C). 
Here, a frequency range of 1320-1620 MHz is shown. The emission is relatively diffuse; no striae are 
obviously present.  The emission is frequency-bounded from above and below throughout the interval. The 
instantaneous frequency bandwidth varies from $\sim 20-180$~MHz ($\Delta\nu/\nu\sim1.4-12.5\%$).  
Fig.~7 also shows a 0.5~s interval and a frequency range of 1120-1220 MHz (interval ``d$_1$" in Fig. 3, panel D). This interval is striking in the abrupt onset and cessation of each patch of emission, and the crisply defined 
lower boundary to the frequency of emission. At least two patches are resolved in frequency, near 
01:15:47.37 and 01:15:47.56 UT. These have bandwidths of 25 and $\sim15$~MHz ($\Delta\nu/\nu\sim1.3$ and 
$2.2\%$, respectively), the latter patch showing signs of band-limited striae. The other patches may 
likewise be composed of overlapping striae drifting from high to low frequency in the same sense as 
displayed in interval ``a".  The upper frequency boundary is not known for the other 
patches because of the presence of the data gap between 1220-1320 MHz, but it is less than 1320~MHz.  
Finally, Fig.~8 again shows 0.5~s of data, but from 1320-1520 MHz, corresponding to the interval ``d$_2$" in 
Fig.~3, panel D. Here, the emission is patchy and diffuse with some striae superposed, again with the 
same sense of drift as noted in prior intervals. Apparent frequency bandwidths 
range from $\sim 5$ MHz to $\gtrsim 200$ MHz ($\Delta\nu/\nu\sim0.5-14\%$).

While these observations exploited the largest frequency-bandwidth ratio currently available (36\%), 
the bandwidth is still 
insufficient to detect possible low-order harmonics of the emission because the ratio of the highest 
to the lowest frequency sampled is only 1.45:1. While higher harmonic ratios could in principle 
be detected -- e.g., 4:3 -- there is no evidence for such harmonic relations in the data. Therefore, 
no constraints on the emission mechanism are possible on the basis of harmonic structure.

To summarize, a variety of spectral structures -- fast-drift striae, discrete patches, and diffuse emission -- 
are seen in the few seconds of data presented here. The striae show a remarkably uniform drift rate of 
$\approx -2.2$ GHz s$^{-1}$. 
The durations of the striae, typically
$\sim$2 ms, imply dynamical source sizes $\approx$0.2\% of the stellar radius.
The spectral features are resolved in frequency, showing bandwidths ranging 
from $\lesssim\!0.5\%$ to as much as 14\%. Where drifting structures appear to be present,
the sense of the drift is the same -- the emission starts at high frequencies and proceeds to lower frequencies.
The signatures appear only in RCP emission, indicating a high degree of circular polarization. 

\section{Discussion}

In this section we focus our attention the striae shown in Fig.~4, the reason being that their properties -- 
flux density, drift rate, mean duration, and mean bandwidth -- are well constrained. We first consider 
whether the observed spectral features -- notably the rapid frequency drift with time -- could result 
from dispersion of the signal in the plasma due to group delay. We also consider propagation effects such as scintillation. We conclude that it is unlikely that propagation effects play a significant role. We then consider the relevant emission mechanism. We argue that, in contrast to the fast drift bursts presented in Paper I, the striae burst in the present case are unlikely the result of plasma radiation from electron 
beams and suggest that they may be due to the action of the cyclotron maser instability (CMI). We comment 
on the other burst phenomena at the end of the section. 

\subsection{Propagation Effects}

As a radio wave traverses a plasma medium, it propagates at the group velocity $v_g=\mu c$, where 
$\mu=\sqrt{1-\nu_{pe}^2/\nu^2}$ is the refractive index and $\nu_{pe}^{2}=e^2 n_e/\pi m_e$ is the electron 
plasma frequency,  where $n_e$ is the electron number density in the plasma medium, $e$ is the electron 
charge, and $m_e$ is the electron mass. If a radio signal emitted at a frequency $\nu$  
is received at a distance $d$ from the source, the delay in reception relative to a signal traversing the same 
distance {in vacuo} is

\begin{equation}
\Delta t = {1\over c} \int_{0}^{d} ({1\over \mu} - 1) dr
\end{equation}

\noindent We first consider the stellar corona and assume that the variation in electron number density can 
be described by a barometric model: $n_e(r)=n_\circ \exp(-r/\lambda_n)$,  
where $\lambda_n$ is the number density scale height. If we assume that the 
observed emission is fundamental plasma radiation emitted at $r=R$, so that $\nu=\nu_{pe}(R)$, then the group 
delay equation can be recast \citep{wild1959}, integrating from the location of the source
emission to $\infty$, as 

\begin{equation}
\Delta t={{2\lambda_n}\over c}\int_0^1{{d\mu}\over{1+\mu}} = {{2\lambda_n\log2}\over{c}}.
\end{equation}

\noindent Under these circumstances the delay is frequency independent and an observed frequency drift 
can be attributed to source motion (Paper I).

On the other hand, if the emission occurs at frequencies well above the local plasma frequency ($\nu\gg\nu_{pe}$), 
then $1/\mu\approx 1+\nu_{pe}^2/2\nu^2$ and the relative delay between two frequencies can be written as\\
\begin{equation}
\Delta t={1\over c}\int_R^\infty \Bigl[{1\over{\mu(\nu_1)}}-{1\over{\mu(\nu_2)}}\Bigr]dr = {{e^2}\over{2\pi m_e c}}\Bigl[{1\over{\nu_1^2}}-{1\over{\nu_2^2}}\Bigr]\int_R^\infty n_e(r)dr
\end{equation} 

\noindent where the integral quantity is commonly referred to as the {\sl dispersion measure}. 
Supposing the source is intrinsically broadband and is located at least $3\lambda$ above the plasma 
level where $n=n_\circ\approx 1.7\times 10^{10}$ cm$^{-3}$, corresponding to a plasma frequency of 
1170 MHz, $\Delta t<1.3 \times 10^{-13}\lambda_{n}$ s. Taking $\lambda_{n}\sim 10^{10}$ cm, $\Delta t< 1.3$ 
ms across the 100 MHz bandwidth, far less than the 45~ms observed in the striae described above.
It is also worth considering the additional 
delay introduced by the interstellar medium. With a mean density of 0.1 cm$^{-3}$ \citep{lism} and a distance to 
AD Leo of 4.9 pc,  we find that the incremental delay resulting from propagation through the ISM is only 0.25~ms 
across the 100 MHz band. 

In fact, rather special conditions are required to account for the observed frequency drift in terms of 
impulsive broadband signals and group delay.  The relative delay between two frequencies $\nu_1$ and $\nu_2$ 
within a source at a height $r$ is \\
\begin{equation}
\Delta t={{2\lambda_{n}}\over c}\int_{\mu_1(r)}^{\mu_2(r)}{{d\mu}\over{1+\mu}} = {{2\lambda_{n}}\over{c}} \log{{1+\mu_2(r)}\over{1+\mu_1(r)}}.
\end{equation}

\noindent where $\mu_{1,2}$ is the refractive index at $r$ for frequencies $\nu_{1,2}$. We find that the 
observed delay of 45~ms between 1120--1220~MHz can occur if the mean frequency of the band is $\approx 20\%$ above the plasma frequency $\nu_{pe}\approx 0.97$~GHz, where $\mu\sim 0.55$. 

For completeness we mention that the presence of a magnetic field renders a plasma birefringent 
and radiation at a fixed frequency shows a differential delay between the ordinary and extraordinary 
modes, corresponding to a delay between RCP and LCP emissions. At frequencies where $\nu\gg \nu_{pe}$ 
and $\nu_{Be}$, the differential group delay between RCP and LCP over a path from the region 
of source emission to observation is \citep{benzpianezzi1997,fleishman2002a,fleishman2002b}\\
\begin{equation}
\Delta t = \int \Bigl[\frac{1}{v_{x}}-\frac{1}{v_{o}}\Bigr] dr ={{2\lambda_{n}}\over c}\  {{\nu_{pe}^2 \nu_{Be}\cos\theta} \over {\nu^3}}
\end{equation}
\noindent where $\nu_{Be}=eB/2\pi m_{e}c$ is the electron gyrofrequency and
$\theta$ is the angle between the magnetic field and the line of sight.  For the 
observations under consideration, the emission is essentially 100\% RCP, discrete LCP features 
being indistinguishable from background. Hence, any differential delay between RCP and LCP is unmeasurable. 
Future observations of moderately polarized bursts with sufficient time resolution and frequency coverage
may be able to exploit this technique.

We now briefly consider additional propagation effects that may influence the observed dynamic spectra. 
Turbulent interstellar plasma can produce scattering of radio waves propagating from point-like sources; for
objects with high enough transverse velocity and/or long path length through the scattering medium,
amplitude modulations, pulse broadening, and frequency smearing can occur.
The slow amplitude variations of pulsars, maser sources, and compact cores of 
active galatic nuclei among other sources, have been identified as refractive interstellar
scintillations (RISS) \citep[see discussion in][]{rickett1990}.
Diffractive interstellar scintillation (DISS) occurs in the 
limit of strong scattering from small-scale irregularities in interstellar plasma,
and corresponds to deep modulations in time and frequency.  
Dynamic spectra of pulsars reveal bands drifting in frequency and time, and criss-cross and periodic 
bands of emission, which qualitatively resemble some of the features seen in AD~Leo's
dynamic spectrum; these features have been explained qualitatively by refractive steering
of a diffractive interstellar scintillation pattern \citep{rickett1990}.
Could the observed spectral structure be due to scintillation as a result of propagation 
through the stellar corona and/or the ISM?
Sources must be compact for either 
phenomenon (RISS or DISS) to occur.  
The small intrinsic source size implied by the extremely short durations of the observed 
emissions ($d < 600$ km) suggests an angular size $\theta\sim d/D \lesssim\!1 \mu$as. 
RISS corresponds to relatively (temporally) slowly varying 
intensities, and also should not change polarization variations \citep{rickett1990}, thus
inconsistent with the emission properties described earlier.
DISS is usually seen in two-dimensional dynamic 
spectra; analysis of the two-dimensional Fourier transform of dynamic spectra reveals discrete 
structures which are typically explained as interference between two or more scattered images. 
Scintillation arcs due to DISS in pulsar dynamic spectra are a broadband phenomenon, having been observed 
over a factor of $\sim$5 in frequency \citep{hill2003}. 
Pulse 
broadening due to multi-path scattering in the ISM can be estimated using equation (6) in 
\citet{cordesmclaughlin2003} and \citet{cordeslazio2001}
which, for a dispersion measure of 0.5 pc cm$^{-3}$, is $\sim$10$^{-10}$ s 
and therefore negligible. The scintillation bandwidth is inversely related to the pulse broadening time, 
which can be estimated as $\Delta \nu_{d}\approx$ 1600 MHz.  Additionally, the scintillation timescale 
can be estimated from the transverse velocity of the star, and is $\sim$5000~s. The time scale of the 
phenomena seen in pulsar dynamic spectra are typically much longer lived than the timescale of the emission 
seen here, as well as being broadband. Thus the characteristics of the observed emission are incompatible with 
RISS and DISS.

Alternatively, the density inhomogeneities in the stellar corona may yield scintillation phenomena. This 
case has been investigated in the case of solar radio emission and coronal turbulence 
\citep{bastian1994,uralov1998}.
However, this possibility can be dismissed by the following argument: 
to observe diffractive scintillation requires enough temporal and spectral resolution to resolve the 
decorrelation time and bandwidth, respectively. The decorrelation time for a point source embedded in the 
corona of AD~Leo is $\tau_d\sim D^2\theta_c^2/2cz\sim d^2/2cz$  
\citep{lee1977}, where $D$ is the distance to 
AD~Leo and $\theta_c\sim d^2/D^2$ is the (angular) source size, $d$ is the (linear) source size estimated 
in \S2, and $z$ is the distance from the source to the effective scattering screen.  \citet{bastian1994} 
showed that $z\sim 0.1$ R$_\odot$ for solar coronal conditions; we assume that it is of similar magnitude 
here in relation to AD~Leo's radius and find that $\tau_d\sim 20$ $\mu$s, far shorter than the 
time resolution of the observations. Moreover, 
the decorrelation bandwidth is $\Delta\nu\sim 1/2\pi\tau_d\sim 20$ kHz, much smaller than the spectral 
resolution of the observations. Hence, diffractive phenomena (scintillations) as a result of scattering 
in the corona of AD~Leo are not expected. 

We conclude from these considerations that the observed spectral structure is intrinsic to the source 
with the possible exception of frequency drift of the striae. However, if the observed drift rate is due to 
the group delay, rather special conditions are required. In particular, the source frequency 
$\nu\approx 1.2\nu_{pe}$ and $\mu\approx 0.55$.  We now discuss the intrinsic spectral structure
and observed properties in light of possible emission mechanisms.

\subsection{Emission Mechanism}

We now consider what emission mechanism is responsible for the radio burst on AD~Leo. We begin with a 
brief comparison of the observational results obtained here with those obtained for the fast-drift 
bursts observed on 2003 June 13, previously described in Paper I. In the case of the bursts on June 13 
they, too, were highly circularly polarized, had frequency drift rates comparable in magnitude to those 
observed for the striae, and had maximum recurrence rates comparable with those found for the striae. 
However, unlike the striae described here, both positive and negative drift rates were observed on 
June 13 in comparable numbers. Moreover, the fast-drift bursts of June 13 were characterized by durations and 
frequency bandwidths each more than an order of magnitude larger -- $\approx\! 30$ ms, and $5\%$, 
respectively -- than the mean values of these parameters obtained for the striae. 
We explored the effect of time resolution on the calculated drift rate by looking at a region
of the dynamic spectrum where several isolated striae occurred, and analyzed the data at 10 ms time resolution
using the method of \citet{ob2006} and the method used here.
For two sub-bursts with measurable drift rates in both sets of data, the differences between the
two methods agreed to within 20\%.  Thus the differences in properties of the fast-drift bursts
observed in 2003 and the striae presented here appear to be intrinsic and may be produced
by a different emission mechanism.

The brightness temperature implied by the burst durations is given by 

\begin{equation}
T_{b} \ge 6\times 10^{14} S_{\rm mJy} \left( \frac{D_{pc}}{\nu_{GHz} \Delta t_{\rm ms}} \right)^{2}
\end{equation}

\noindent where S$_{\rm mJy}$ is the source flux in mJy, $D_{pc}$ is the source distance in pc, $\nu_{GHz}$ 
is the central frequency, $\Delta t_{\rm ms}$ is the light travel time across the source in ms, taken 
to be comparable to the burst duration. In the case of the June 13 fast-drift bursts, brightness 
temperature limits of $T_B\gtrsim 4\times 10^{14}$ K were obtained whereas with mean durations of only 2~ms 
and with S$_{\rm mJy}\approx$500, $\nu_{GHz}\sim$1.2, $T_B \gtrsim 10^{18}$K for the striae. This places 
rather stringent constraints on the source emission mechanism. 

Paper I demonstrated that the fast drift bursts shared attributes with solar decimetric spike 
bursts \citep{gudelbenz1990} and concluded that they could be the result of (ordinary mode) plasma 
radiation near the fundamental of the local electron plasma frequency, 
driven by multitudes of 
suprathermal electron beams. Consistent with this idea is the observed burst durations: if the duration of the 
emission is determined by collisional damping, a coronal plasma temperature of 13 MK is implied, 
comparable with the dominant temperature inferred from soft X-ray observations \citep{maggio2004}.

The small instantaneous bandwidth of the striae observed here implies that if fundamental plasma radiation is the relevant emission 
mechanism, Langmuir waves are excited over a small range of densities at any given time. In particular, $\Delta n_e/n_e = 2\Delta\nu/\nu\lesssim 
1\%$; then if $n_e=n_\circ exp(-r/\lambda_n)$, we have $\Delta n_e/n_e\approx \delta s/\lambda_n$, or $\delta s\lesssim 0.01\lambda_n$, 
where $\delta s$ is the extent of the exciter along its trajectory. Again, with $\lambda_n\sim 10^{10}$ cm, $\delta s\lesssim 1000$ km, 
consistent with previous estimates of the source size. Moreover, if the duration of discrete striae is determined by collisional damping, a 
relatively cool source is implied, only 1.5 MK for fundamental plasma radiation. While relatively cool plasma does exist in the corona of AD~Leo 
\citep{vdb2003} the assumption that the emission is fundamental plasma radiation implies a relatively dense source with $n_{10}\sim 
2$. The picture to emerge is one in which the striae are excited by a succession of tiny blobs, each a few hundred km in size, in a relatively 
cool, dense plasma. A significant problem with this idea is that if the source were in such a cool, dense plasma, the optical depth $\tau$ 
of the overlying plasma due to collisional absorption would be extremely large, with $\tau\sim 300 (\lambda_n/10^{10})$  \citep[e.g.,][]{benz2002}. 
In this case, it would be difficult to understand the escape of the radiation at all, let alone the extremely high brightness temperature 
of the bursts, unless $\lambda_n$ is at least two orders of magnitude smaller than assumed, as perhaps might be possible in a highly 
inhomogeneous corona. But this idea cannot be correct. The striae are observed to drift in frequency at a rate of 2.2 GHz s$^{-1}$. The 
drift rate is $\dot\nu=\nu_{pe} v_b/2\lambda_n\approx \nu v_b/2\lambda_n$, where $v_b$ is the speed of the blobs presumed to excite Langmuir 
waves and $\nu\approx \nu_{pe}$. Then if $\lambda_n\sim 10^8$ cm, we infer that $v_b \sim 3800$ km s$^{-1}$, which is comparable to the thermal 
speed of the ambient electrons with $T\sim 1.5$ MK and an instability that produces the necessary Langmuir waves is not expected.

On these grounds, therefore, we question the relevance of plasma radiation to the striae described here. 
What is the alternative? One possibility is clearly the cyclotron maser instability (CMI), a mechanism 
widely believed to play a significant role in planetary magnetospheres 
\citep[see the recent review by][and references therein]{treumann2006}
and long suspected of playing a role in the radio emission from the 
Sun \citep{melrosedulk1982}, stellar coronae \citep[e.g.,][]{bb1987,gudel1989,bastian1990,abada1994,bingham2001}
and, more recently, in the coronae of 
extremely late-type stellar and substellar objects \citep[e.g.,][]{hallinan2007}. The CMI is a resonant 
wave-particle phenomenon, the resonance being between electromagnetic waves and magnetized electrons. The 
source of free energy for the process is an anisotropy in the electron distribution function that produces a 
positive gradient with respect to the perpendicular momentum: $\partial f/\partial p_\perp > 0$. In its 
simplest form, the anisotropy is assumed to take the form of a loss-cone distribution, as might be set up in 
closed coronal magnetic loops. However, work over many years in the context of terrestrial and planetary 
radio emissions has greatly extended and refined our knowledge about the conditions under which the CMI is 
operative. \citet{treumann2006} reviews the extensive {\sl in situ} observations that have been made in the 
terrestrial magnetosphere over the past two decades. These demonstrate that loss cone anisotropies are 
not the dominant driver of the CMI; instead ``shell" or ``horseshoe" distributions are relevant, with 
magnetic field aligned electric fields playing a central role in setting up the distribution that 
provides the free energy to the CMI \citep[e.g.,][]{ergun2000}. Moreover, the CMI occurs in density 
cavities where $\nu_{pe}\ll \nu_{\rm Be}$. Here, $\nu_{Be}=eB/2\pi m_e c\approx 2.8 B_3$ GHz 
and  $B_3$ is the magnitude of the magnetic field in kG. We note that if the 
CMI is the relevant mechanism, with $\nu_{pe}\ll \nu_{Be}$, group delay is insignificant and the 
observed drifts are intrinsic to the source.

The CMI leads to direct amplification of electromagnetic waves, producing intense, coherent radiation 
at the fundamental or possibly, the harmonic, of the (relativistic) electron gyrofrequency $\nu_{\rm Be}/\gamma$. 
The radiation is emitted in a narrow angular range perpendicular to the magnetic field for horseshoe 
distributions. The extraordinary (x) mode is generally favored over the ordinary (o) mode for growth by the 
CMI, thereby explaining the high degree of circular polarization. Its frequency bandwidth is expected to 
be of order $1\%$ or less, although broader frequency bandwidths can be accomodated with higher energy electrons. 
It is expected to achieve brightness temperatures in excess of $10^{18}$ K. 
These attributes are collectively consistent with the observations of the striae.  

What about the drift rate of the striae? In recent years, there has been great interest in fine structure 
seen in CMI emission in the terrestrial case, driving speculation concerning the role of  ``elementary 
radiation sources". Perhaps analogous to the striae seen on AD Leo is so-called striped or striated 
terrestrial auroral kilometric radiation \citep[AKR; see, e.g., examples presented by][]{menietti1996,menietti2000,
pottelette2001,mutel2006}.
\citeauthor{pottelette2001} argue that such fine structure may be due to 
localized structures -- electron or ion phase space holes -- in the source region excited by the 
counter-streaming electrons and ions. The steep gradients set up in the local distribution function by 
these holes may contribute intense, elementary sources of radiation \citep[see][for a detailed discussion]{treumann2006}.
The frequency drift of AKR fine structures is interpreted as the propagation speed of these 
elementary radiation sources along the magnetic field in the source. Speeds of order 500 km s$^{-1}$ are 
inferred for the case of AKR fine structures. Applying this to the case of the striae on AD~Leo, the drift rate is 
presumed to be intrinsic to the source and therefore represents motion of the source along the magnetic 
field gradient: $\dot\nu\propto \dot B = v\nabla B$. If the magnetic field can be described as 
dipolar, $B=B_\circ (\lambda_B/r)^3$, where $\lambda_B$ is the characteristic scale of the dipole 
field, and the observed frequency is the local electron gyrofrequency, the speed of the exciter can be written $v\approx 47 \lambda_B B_\circ^{1/3} \dot\nu \nu^{-4/3}$. 
With $\nu$ and $\dot\nu$ known, this becomes $v\approx 0.084\lambda_B B_\circ^{1/3}$. 
Assuming that each striae represents an elementary radiation source, capping the 
speed of these sources to be less than that of the emitting electrons ($\sim 10$ keV) 
suggests that $v<6\times 10^4$ km s$^{-1}$ and so $(\lambda_B/r_\star) B_\circ^{1/3}<2.4$, where $r_\star$ 
is the radius of AD~Leo (0.4 R$_\odot$). For $B_\circ = 500-2000$ G, we find that $\lambda_B 
\lesssim 0.2-0.3 r_\star$. 
We conclude that the fast-drift striae may be compatible with the 
CMI mechanism if the magnetic field in the source is described by a magnetic field with scale 
comparable to a large ``active region" rather than a global dipole field.

If the CMI mechanism is indeed relevant, the problem first pointed out by \citet{melrosedulk1982} and 
reiterated in Paper I remains: how does the radiation escape from the source to a distant observer. 
CMI radiation emitted near the fundamental of the electron gyrofrequency suffers catastrophic absorption 
at the second gyroresonant harmonic layer of the atmosphere. \citet{ergun2000} argue, however, that because the 
CMI operates in a density cavity in the terrestrial case, the surrounding plasma acts as a duct. Rapid 
refraction and scattering of CMI radiation cause it to emerge more nearly parallel to the magnetic field. Alternatively, \citet{robinson1989}
has suggested that conditions may be favorable for partial mode conversion in CMI sources. In 
particular, fundamental x-mode radiation amplified by the CMI undergoes partial conversion to 
fundamental o-mode. The optical depth to o-mode radiation can be several hundred times less 
than that to x-mode. This process does not require significant scattering or refraction of the 
emitted radiation.  
Whether either of these processes in fact occurs in the corona of AD~Leo cannot be answered by the 
observations in the present case.

\section{Concluding remarks}

We have described observations of a unique set of stellar radio bursts, which 
take advantage of the wide bandwidth and high time resolution
capabilities of the Wideband Arecibo Pulsar Processor at the Arecibo Observatory.  
These ultra-high time resolution observations reveal phenomena that differ from those previously
described using a similar observational setup, pointing out the
complexity and diversity of processes likely occurring in stellar
coronal plasmas.  Whereas in Paper I we concluded that a plasma
emission process appeared to be producing the two types of radio
bursts observed in June 2003, in the current paper we prefer a different explanation, a cyclotron
maser instability, for the
fast-drift striae observed in April 2005.  While all sets of phenomena show drifting
structures of highly circularly polarized radiation, 
key discriminants between them 
are the durations and 
bandwidths of spectral features, as well as the magnitude and sign of the drift rates.  

In Paper I and here, we have demonstrated that the analysis of dynamic spectra of stellar radio bursts 
provide observational constraints which can be used as a measure
to gauge the likelihood that a particular emission process is
operative.
Extensions of the current observational setup can look for
dynamics at even higher time resolution, search for harmonic emissions
over larger frequency bandwidths, expand the observational program
to other dMe flare stars, and search for high time resolution behavior on 
other classes of active stars.
Given the complexity of solar radio emissions at meter wavelengths compared with the
already rich variety of decimetric phenomena, the observational results presented
here for the dMe flare star AD Leo suggest that the next generation of radio instrumentation,
particularly at metric wavelengths, promises to reveal a wealth of new 
phenomena which can diagnose plasma processes occurring in stellar coronae.
As highly circularly polarized radio emission appears to be a common phenomenon on 
active stars, 
these spectacular radio bursts on M dwarf flare stars apparently represent the tip 
of the iceberg of stellar coronal plasma physics soon to be available for study.

\acknowledgements{We thank Phil Perillat and Avinash Deshpande at NAIC for their outstanding assistance in making this observational program a success.  We thank the referee, Manuel G\"{u}del, for a careful reading of the paper.
This paper represents the results of program A2013 at Arecibo Observatory.}


\begin{thebibliography}{41}
\expandafter\ifx\csname natexlab\endcsname\relax\def\natexlab#1{#1}\fi

\bibitem[{{Abada-Simon} {et~al.}(1997){Abada-Simon}, {Lecacheux}, {Aubier}, \&
  {Bookbinder}}]{abada1997a}
{Abada-Simon}, M., {Lecacheux}, A., {Aubier}, M., \& {Bookbinder}, J.~A. 1997,
  \aap, 321, 841

\bibitem[{{Abada-Simon} {et~al.}(1994){Abada-Simon}, {Lecacheux}, {Louarn},
  {Dulk}, {Belkora}, {Bookbinder}, \& {Rosolen}}]{abada1994}
{Abada-Simon}, M., {Lecacheux}, A., {Louarn}, P., {Dulk}, G.~A., {Belkora}, L.,
  {Bookbinder}, J.~A., \& {Rosolen}, C. 1994, \aap, 288, 219

\bibitem[{{Bastian}(1994)}]{bastian1994}
{Bastian}, T.~S. 1994, \apj, 426, 774

\bibitem[{{Bastian} {et~al.}(1990){Bastian}, {Bookbinder}, {Dulk}, \&
  {Davis}}]{bastian1990}
{Bastian}, T.~S., {Bookbinder}, J., {Dulk}, G.~A., \& {Davis}, M. 1990, \apj,
  353, 265

\bibitem[{{Bastian} \& {Bookbinder}(1987)}]{bb1987}
{Bastian}, T.~S. \& {Bookbinder}, J.~A. 1987, \nat, 326, 678

\bibitem[{{Benz}(2002)}]{benz2002}
{Benz}, A., ed. 2002, Astrophysics and Space Science Library, Vol. 279, {Plasma
  Astrophysics, second edition}

\bibitem[{{Benz} \& {Pianezzi}(1997)}]{benzpianezzi1997}
{Benz}, A.~O. \& {Pianezzi}, P. 1997, \aap, 323, 250

\bibitem[{{Bingham} {et~al.}(2001){Bingham}, {Cairns}, \&
  {Kellett}}]{bingham2001}
{Bingham}, R., {Cairns}, R.~A., \& {Kellett}, B.~J. 2001, \aap, 370, 1000

\bibitem[{{Cordes} \& {Lazio}(2001)}]{cordeslazio2001}
{Cordes}, J.~M. \& {Lazio}, T.~J.~W. 2001, \apj, 549, 997

\bibitem[{{Cordes} \& {McLaughlin}(2003)}]{cordesmclaughlin2003}
{Cordes}, J.~M. \& {McLaughlin}, M.~A. 2003, \apj, 596, 1142


\bibitem[{{Cox} \& {Reynolds}(1987)}]{lism}
{Cox}, D.~P. \& {Reynolds}, R.~J. 1987, \araa, 25, 303

\bibitem[{{Ergun} {et~al.}(2000){Ergun}, {Carlson}, {McFadden}, {Delory},
  {Strangeway}, \& {Pritchett}}]{ergun2000}
{Ergun}, R.~E., {Carlson}, C.~W., {McFadden}, J.~P., {Delory}, G.~T.,
  {Strangeway}, R.~J., \& {Pritchett}, P.~L. 2000, \apj, 538, 456

\bibitem[{{Favata} {et~al.}(2000){Favata}, {Micela}, \&
  {Reale}}]{favataetal2000}
{Favata}, F., {Micela}, G., \& {Reale}, F. 2000, \aap, 354, 1021

\bibitem[{{Fleishman} {et~al.}(2002{\natexlab{a}}){Fleishman}, {Fu}, {Huang},
  {Melnikov}, \& {Wang}}]{fleishman2002a}
{Fleishman}, G.~D., {Fu}, Q.~J., {Huang}, G.-L., {Melnikov}, V.~F., \& {Wang},
  M. 2002{\natexlab{a}}, \aap, 385, 671

\bibitem[{{Fleishman} {et~al.}(2002{\natexlab{b}}){Fleishman}, {Fu}, {Wang},
  {Huang}, \& {Melnikov}}]{fleishman2002b}
{Fleishman}, G.~D., {Fu}, Q.~J., {Wang}, M., {Huang}, G.-L., \& {Melnikov},
  V.~F. 2002{\natexlab{b}}, Physical Review Letters, 88, 251101

\bibitem[{{G\"{u}del} \& {Benz}(1990)}]{gudelbenz1990}
{G\"{u}del}, M. \& {Benz}, A.~O. 1990, \aap, 231, 202

\bibitem[{{G\"{u}del} {et~al.}(1989){G\"{u}del}, {Benz}, {Bastian}, {Furst},
  {Simnett}, \& {Davis}}]{gudel1989}
{G\"{u}del}, M., {Benz}, A.~O., {Bastian}, T.~S., {Furst}, E., {Simnett},
  G.~M., \& {Davis}, R.~J. 1989, \aap, 220, L5

\bibitem[{{Hallinan} {et~al.}(2007){Hallinan}, {Bourke}, {Lane}, {Antonova},
  {Zavala}, {Brisken}, {Boyle}, {Vrba}, {Doyle}, \& {Golden}}]{hallinan2007}
{Hallinan}, G., {Bourke}, S., {Lane}, C., {Antonova}, A., {Zavala}, R.~T.,
  {Brisken}, W.~F., {Boyle}, R.~P., {Vrba}, F.~J., {Doyle}, J.~G., \& {Golden},
  A. 2007, \apjl, 663, L25

\bibitem[{{Hawley} {et~al.}(2003){Hawley}, {Allred}, {Johns-Krull}, {Fisher},
  {Abbett}, {Alekseev}, {Avgoloupis}, {Deustua}, {Gunn}, {Seiradakis}, {Sirk},
  \& {Valenti}}]{hawleyetal2003}
{Hawley}, S.~L., {Allred}, J.~C., {Johns-Krull}, C.~M., {Fisher}, G.~H.,
  {Abbett}, W.~P., {Alekseev}, I., {Avgoloupis}, S.~I., {Deustua}, S.~E.,
  {Gunn}, A., {Seiradakis}, J.~H., {Sirk}, M.~M., \& {Valenti}, J.~A. 2003,
  \apj, 597, 535

\bibitem[{{Hawley} {et~al.}(1995){Hawley}, {Fisher}, {Simon}, {Cully},
  {Deustua}, {Jablonski}, {Johns-Krull}, {Pettersen}, {Smith}, {Spiesman}, \&
  {Valenti}}]{hawleyetal1995}
{Hawley}, S.~L., {Fisher}, G.~H., {Simon}, T., {Cully}, S.~L., {Deustua},
  S.~E., {Jablonski}, M., {Johns-Krull}, C.~M., {Pettersen}, B.~R., {Smith},
  V., {Spiesman}, W.~J., \& {Valenti}, J. 1995, \apj, 453, 464

\bibitem[{{Hawley} \& {Pettersen}(1991)}]{hawleypettersen1991}
{Hawley}, S.~L. \& {Pettersen}, B.~R. 1991, \apj, 378, 725

\bibitem[{{Hill} {et~al.}(2003){Hill}, {Stinebring}, {Barnor}, {Berwick}, \&
  {Webber}}]{hill2003}
{Hill}, A.~S., {Stinebring}, D.~R., {Barnor}, H.~A., {Berwick}, D.~E., \&
  {Webber}, A.~B. 2003, \apj, 599, 457

\bibitem[{{H{\"u}nsch} {et~al.}(1999){H{\"u}nsch}, {Schmitt}, {Sterzik}, \&
  {Voges}}]{hunschetal1999}
{H{\"u}nsch}, M., {Schmitt}, J.~H.~M.~M., {Sterzik}, M.~F., \& {Voges}, W.
  1999, \aaps, 135, 319

\bibitem[{{Jackson} {et~al.}(1989){Jackson}, {Kundu}, \& {White}}]{jackson1989}
{Jackson}, P.~D., {Kundu}, M.~R., \& {White}, S.~M. 1989, \aap, 210, 284

\bibitem[{{Lee}(1977)}]{lee1977}
{Lee}, L.~C. 1977, \apj, 218, 468

\bibitem[{{Maggio} {et~al.}(2004){Maggio}, {Drake}, {Kashyap}, {Harnden},
  {Micela}, {Peres}, \& {Sciortino}}]{maggio2004}
{Maggio}, A., {Drake}, J.~J., {Kashyap}, V., {Harnden}, Jr., F.~R., {Micela},
  G., {Peres}, G., \& {Sciortino}, S. 2004, \apj, 613, 548

\bibitem[{{McLean} \& {Labrum}(1985)}]{1985srph}
{McLean}, D.~J. \& {Labrum}, N.~R. 1985, {Solar radiophysics: Studies of
  emission from the sun at metre wavelengths} (Solar Radiophysics: Studies of
  Emission from the Sun at Metre Wavelengths)

\bibitem[{{Melrose} \& {Dulk}(1982)}]{melrosedulk1982}
{Melrose}, D.~B. \& {Dulk}, G.~A. 1982, \apj, 259, 844

\bibitem[{{Menietti} {et~al.}(2000){Menietti}, {Persoon}, {Pickett}, \&
  {Gurnett}}]{menietti2000}
{Menietti}, D., {Persoon}, A., {Pickett}, J., \& {Gurnett}, D. 2000, Journal of
  Geophysical Research (Space Physics), 105, 18857

\bibitem[{{Menietti} {et~al.}(1996){Menietti}, {Wong}, {Kurth}, {Gurnett},
  {Granroth}, \& {Groene}}]{menietti1996}
{Menietti}, J.~D., {Wong}, H.~K., {Kurth}, W.~S., {Gurnett}, D.~A., {Granroth},
  L.~J., \& {Groene}, J.~B. 1996, \jgr, 101, 10673

\bibitem[{{Mutel} {et~al.}(2006){Mutel}, {Menietti}, {Christopher}, {Gurnett},
  \& {Cook}}]{mutel2006}
{Mutel}, R.~L., {Menietti}, J.~D., {Christopher}, I.~W., {Gurnett}, D.~A., \&
  {Cook}, J.~M. 2006, Journal of Geophysical Research (Space Physics), 111,
  10203

\bibitem[{{Osten} \& {Bastian}(2006)}]{ob2006}
{Osten}, R.~A. \& {Bastian}, T.~S. 2006, \apj, 637, 1016 (Paper I)

\bibitem[{{Pottelette} {et~al.}(2001){Pottelette}, {Treumann}, \&
  {Berthomier}}]{pottelette2001}
{Pottelette}, R., {Treumann}, R.~A., \& {Berthomier}, M. 2001, \jgr, 106, 8465

\bibitem[{{Rickett}(1990)}]{rickett1990}
{Rickett}, B.~J. 1990, \araa, 28, 561

\bibitem[{{Robinson}(1989)}]{robinson1989}
{Robinson}, P.~A. 1989, \apjl, 341, L99

\bibitem[{{Seiradakis} {et~al.}(1995){Seiradakis}, {Avgoloupis}, {Mavridis},
  {Varvoglis}, \& {Fuerst}}]{seiradakis1995}
{Seiradakis}, J.~H., {Avgoloupis}, S., {Mavridis}, L.~N., {Varvoglis}, P., \&
  {Fuerst}, E. 1995, \aap, 295, 123

\bibitem[{{Stepanov} {et~al.}(2001){Stepanov}, {Kliem}, {Zaitsev}, {F{\"u}rst},
  {Jessner}, {Kr{\"u}ger}, {Hildebrandt}, \& {Schmitt}}]{stepanov2001}
{Stepanov}, A.~V., {Kliem}, B., {Zaitsev}, V.~V., {F{\"u}rst}, E., {Jessner},
  A., {Kr{\"u}ger}, A., {Hildebrandt}, J., \& {Schmitt}, J.~H.~M.~M. 2001,
  \aap, 374, 1072

\bibitem[{{Treumann}(2006)}]{treumann2006}
{Treumann}, R.~A. 2006, \aapr, 13, 229

\bibitem[{{Uralov}(1998)}]{uralov1998}
{Uralov}, A.~M. 1998, \solphys, 183, 133

\bibitem[{{van den Besselaar} {et.~al.}(2003){van den Besselaar}, {Raassen}, {Mewe},
{van der Meer}, {G\"{u}del}, \& {Audard}}]{vdb2003}
{van den Besselaar}, E.~J.~M., {Raassen}, A.~J.~J., {Mewe}, R., {van der Meer}, R.~L.~J.,
{G\"{u}del}, M., \& Audard, M. 2003 \aap, 411, 587 

\bibitem[{{Wild} {et~al.}(1959){Wild}, {Sheridan}, \& {Neylan}}]{wild1959}
{Wild}, J.~P., {Sheridan}, K.~V., \& {Neylan}, A.~A. 1959, Australian Journal
  of Physics, 12, 369

\bibitem[{{Zaitsev} {et~al.}(2004){Zaitsev}, {Kislyakov}, {Stepanov}, {Kliem},
  \& {Furst}}]{zaitsev2004}
{Zaitsev}, V.~V., {Kislyakov}, A.~G., {Stepanov}, A.~V., {Kliem}, B., \&
  {Furst}, E. 2004, Astronomy Letters, 30, 319

\end{thebibliography}

\begin{figure}
\includegraphics[scale=0.7]{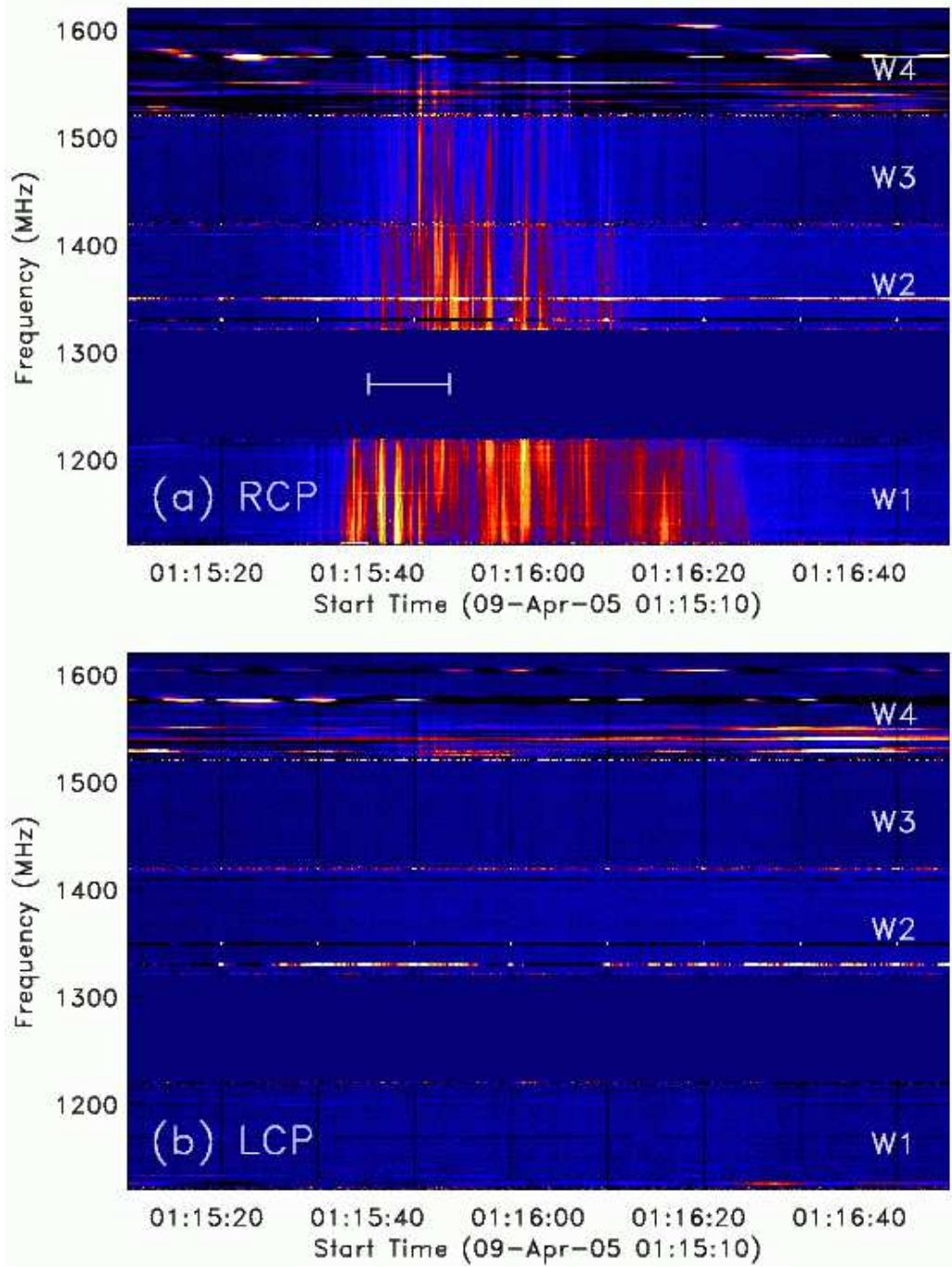}
\caption{Overview of the radio burst observed on AD Leo. a) Right-circularly polarized emission. b) 
Left-circularly polarized emission. The data have been averaged to 100 ms time resolution. The four 
WAPP data channels are labeled W1, W2, W3, and W4. The frequency range of 1220-1320 MHz was deliberately 
avoided due to the presence of strong RFI. Nevertheless, bright and dark horizontal bands and faint periodic vertical bands indicate the presence of RFI that could not be avoided. A detail of the 10~s time interval indicated by the horizontal bar is shown in Fig.~2.
\label{apr9fig}}
\end{figure}

\begin{figure}
\includegraphics[scale=0.85]{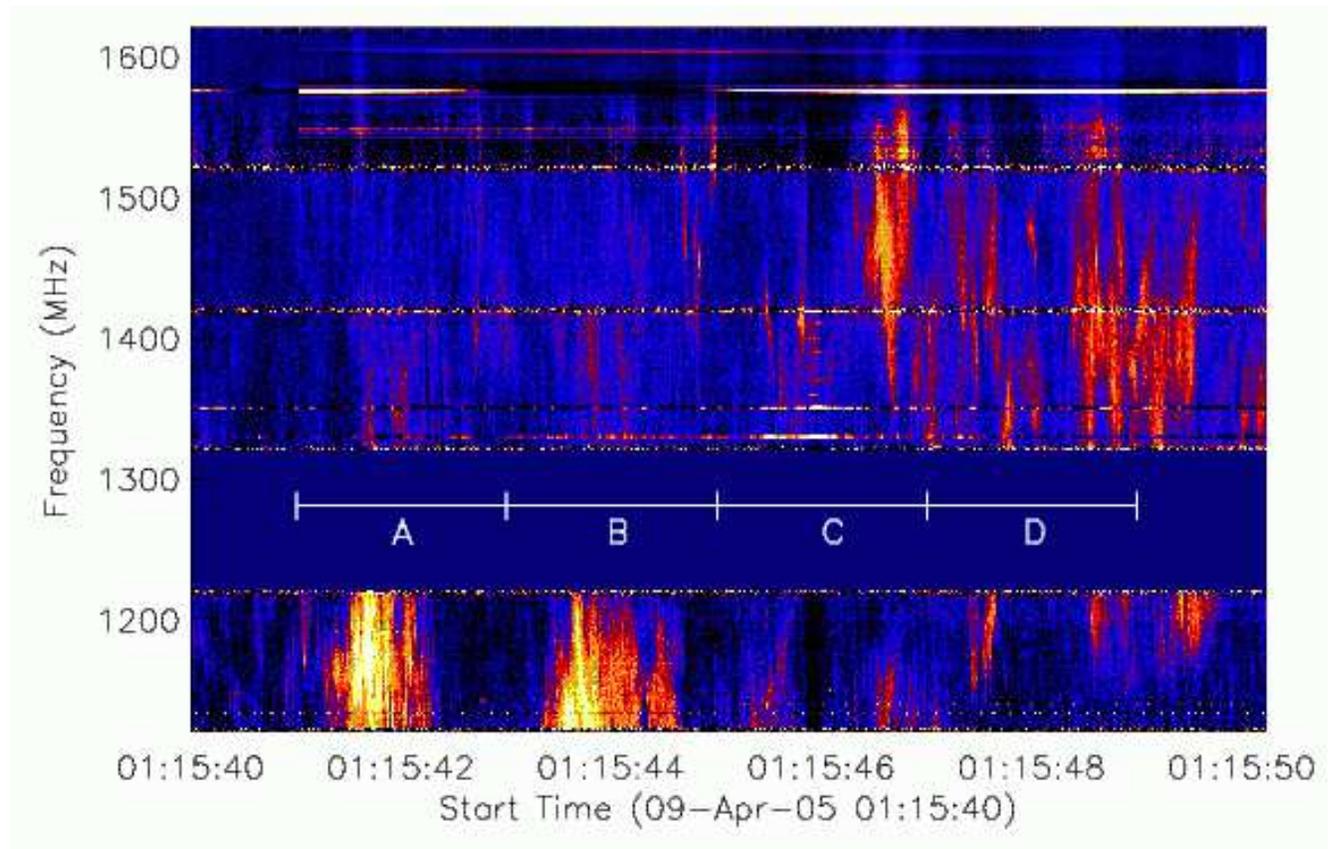}
\caption{A detail of the RCP spectrum indicated by the bar in Fig.~1a at a time resolution of 10~ms. Further details of the spectrum for consecutive 2~s time intervals labeled A, B, C, and D are shown in Fig.~3.
\label{rgn1_dspec}}
\end{figure}

\begin{figure}
\includegraphics[scale=0.7]{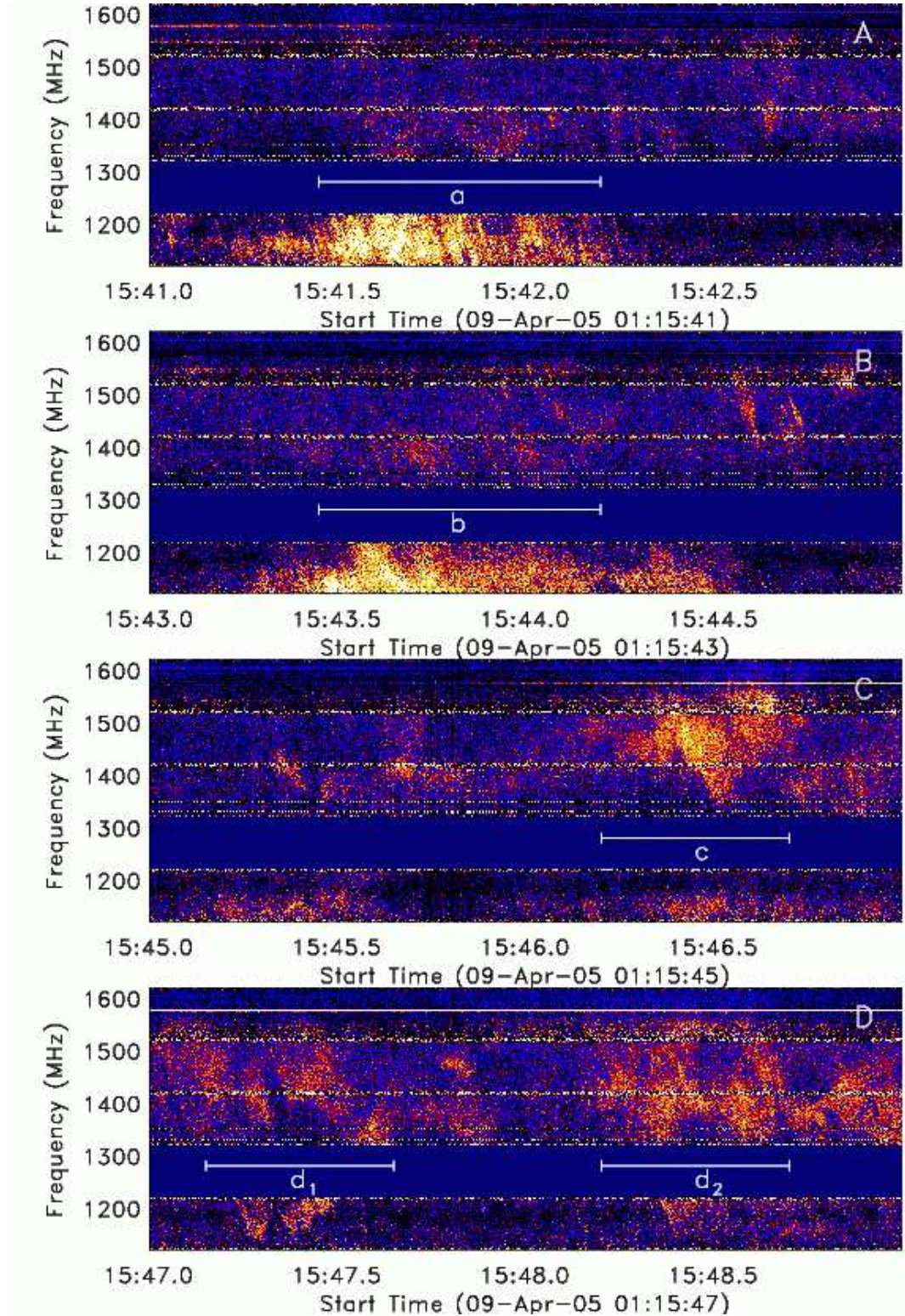}
\caption{Details of the RCP spectrum shown in Fig.~2. Each panel A-D shows 2~s of data with the full time resolution of 1~ms.}
\end{figure}

\begin{figure}
\includegraphics[]{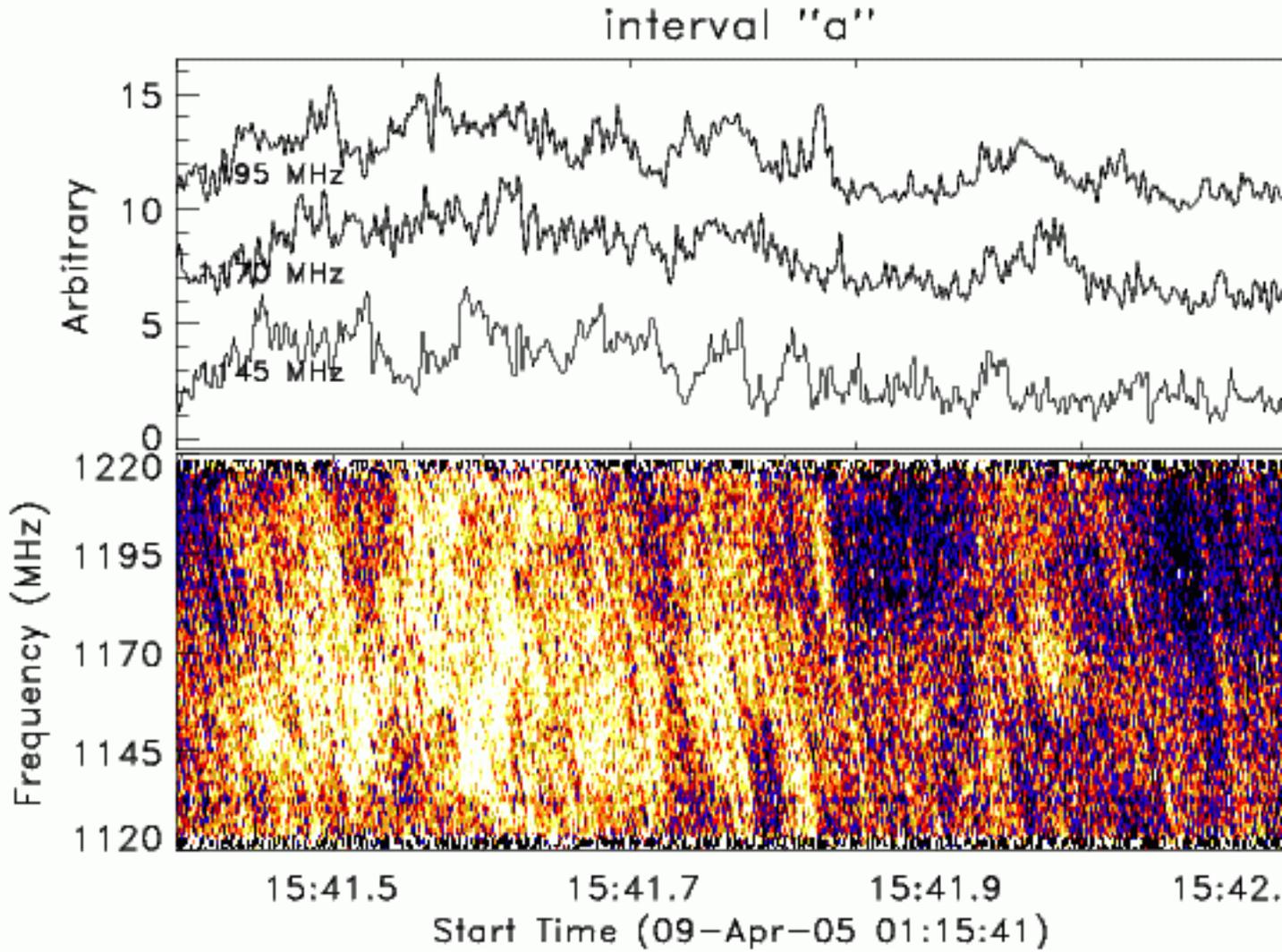}
\caption{Detail of the RCP spectrum indicated by the bar in Fig.~3, panel A, labeled ``a". The frequency range shown is 1120-1220 MHz, the time interval is 0.75~s, and the time resolution is 1~ms. The upper panel shows the time variation of flux density at the frequencies indicated. Note the fast-drift striations.}
\end{figure}

\begin{figure}
\includegraphics[]{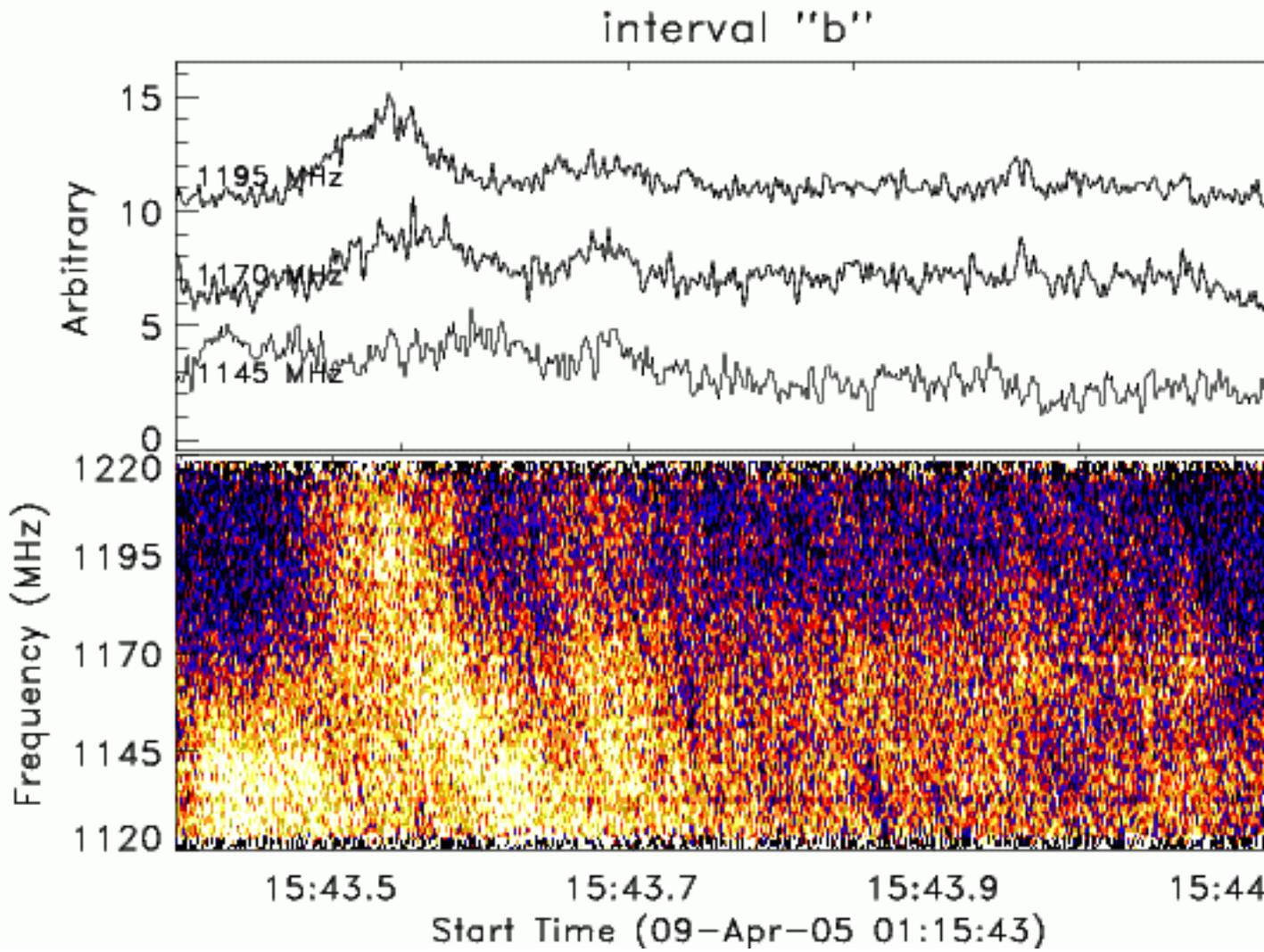}
\caption{Detail of the RCP spectrum as indicated by the bar in Fig.~3, panel B, labeled ``b". The frequency range shown is 1120-1220 MHz, the time interval is 0.75~s,  and the time resolution is 1~ms. Here the emission is relatively diffuse.}
\end{figure}

\begin{figure}
\includegraphics[scale=0.8]{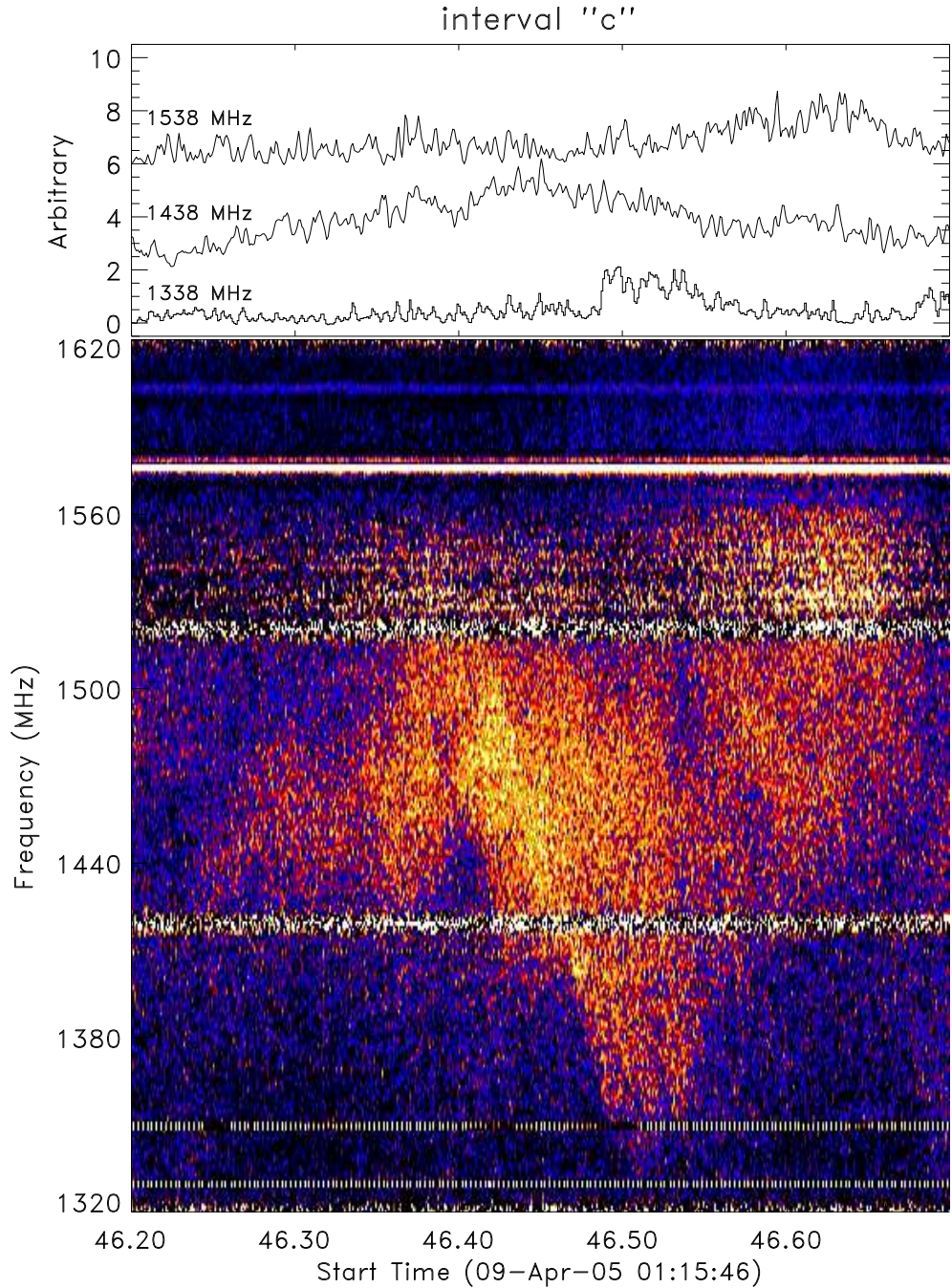}
\caption{Detail of the RCP spectrum as indicated by the bar in Fig.~3, panel C, labeled ``c". The frequency range shown is 1320-1620 MHz, the time interval is 0.5~s, and the time resolution is 1~ms. The emission is again relatively diffuse, with considerable variation in bandwidth.}
\end{figure}

\begin{figure}
\includegraphics[]{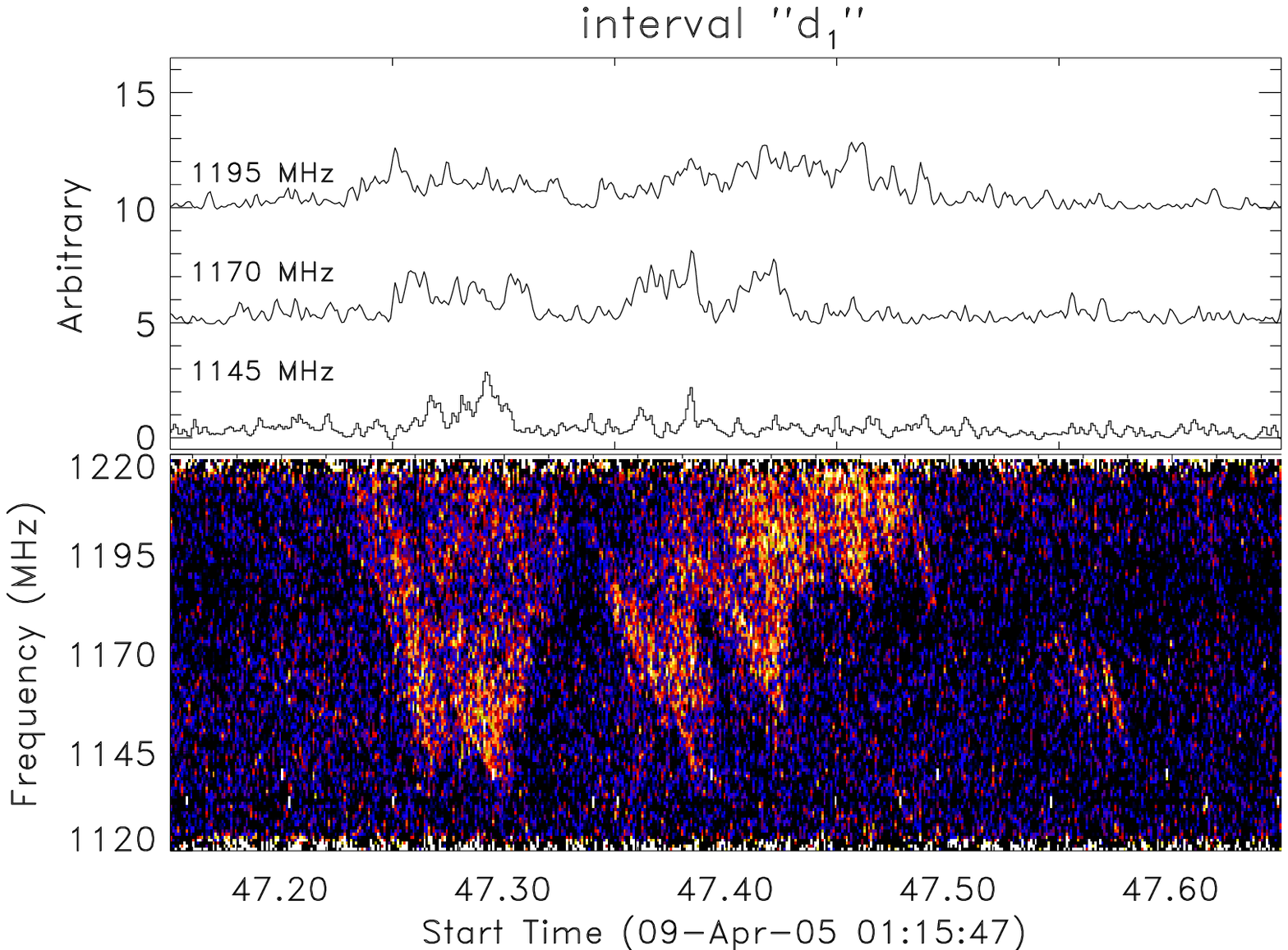}
\caption{Detail of the RCP spectrum as indicated by the bar in Fig.~3, panel D, labeled ``d$_1$". The frequency range shown is 1120-1220 MHz, the time interval is 0.5~s, and the time resolution is 1~ms. }
\end{figure}

\begin{figure}
\includegraphics[]{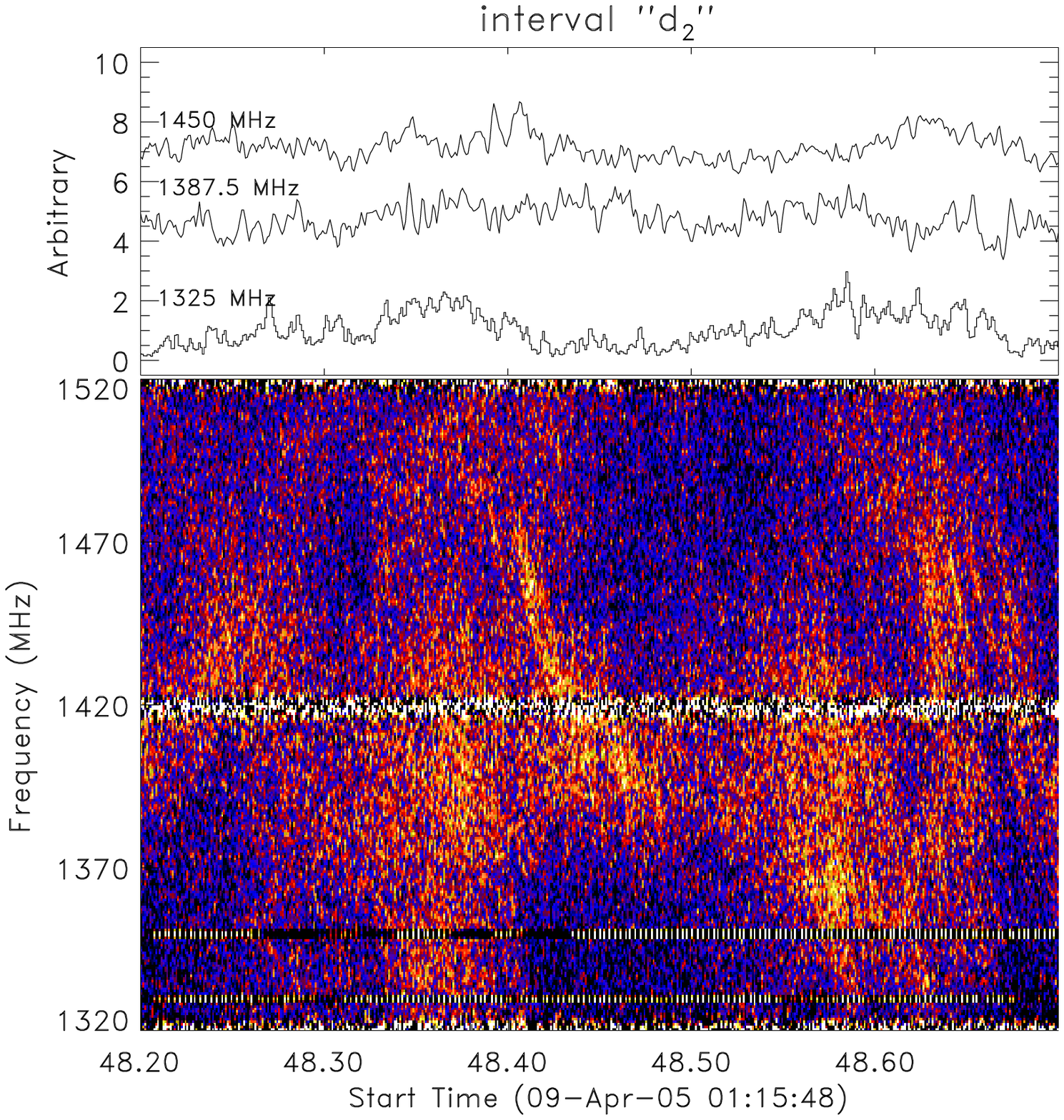}
\caption{Detail of the RCP spectrum as indicated by the bar in Fig.~3, panel D, labeled ``d$_2$". The frequency range shown is 1320-1520 MHz, the time interval is 0.5~s, and the time resolution is 1~ms. }
\end{figure}

\end{document}